\documentclass[%
amsmath,amssymb,
aps,
pra,
]{revtex4-2}

\usepackage{hyperref}

\usepackage{graphicx}% Include figure files
\usepackage{subcaption}
\usepackage{dcolumn}% Align table columns on decimal point
\usepackage{bm}% bold math
\usepackage{amsmath, amsthm, amssymb}
\usepackage{mathtools}

\usepackage{hhline}
\usepackage{nicefrac}

\usepackage{footnote}
\usepackage{xcolor}

\newcommand{\dd}{{\rm d}}

\usepackage[mathscr]{eucal}

\usepackage{color}
\usepackage{subcaption}

\begin{document}

\def\be{\begin{equation}}
\def\ee{\end{equation}}

\title{Born rule in Schr\"{o}dinger--Newton scenarios of semiclassical gravity}

\author{Tanguy Fontana}
\author{Axelle Coulon}

\author{Giovanni Manfredi}
\email{giovanni.manfredi@ipcms.unistra.fr}
\affiliation{Universit\'e de Strasbourg, CNRS, Institut de Physique et Chimie des Mat\'eriaux de Strasbourg, UMR 7504, F-67000 Strasbourg, France}

\begin{abstract}
We investigate dynamical relaxation to the Born rule in a semiclassical gravity model described by the Schr\"odinger--Newton equations. Using a Gaussian variational solution for the self-gravitating wave packet, combined with Nelson’s stochastic representation of quantum dynamics, we track the evolution of particle ensembles whose initial probability distribution differs from $|\Psi|^2$. Relaxation to the Born rule (``quantum equilibrium") occurs when the wave packet remains bounded, or grows sufficiently slowly in time; in contrast, for rapidly expanding solutions of the Schr\"odinger--Newton equations, deviations from the Born rule can persist indefinitely. Numerical simulations confirm this analytical picture. 
These results show that semiclassical self-gravity can either enable or inhibit relaxation towards the Born rule, providing a minimal model relevant to proposals of primordial quantum non-equilibrium, as suggested by Underwood and Valentini [Phys. Rev. D {\bf 92}, 063531 (2015)].
\end{abstract}

\maketitle

\section{Introduction}\label{sec:intro}

% Review of SNEs
\subsection{Schrödinger--Newton equations}

The Schrödinger--Newton equations (SNEs),
\begin{eqnarray}
i\hbar\,\frac{\partial\Psi}{\partial t}
&=& -\frac{\hbar^{2}}{2m}\,\Delta\Psi + m\,V(\mathbf r,t)\,\Psi, \label{eq:schrodinger}
\\
\Delta V &=& 4\pi G\,m\,|\Psi|^{2}, \label{eq:newton}
\end{eqnarray}
constitute the simplest nonrelativistic framework in which quantum dynamics is coupled to Newtonian gravity. Here $\Psi(\mathbf r,t)$ denotes the wavefunction, $V(\mathbf r,t)$ the Newtonian gravitational potential, $m$ the particle mass, $G$ the gravitational constant, and $\hbar$ Planck’s constant. The system is intrinsically nonlinear, since the mass density $\rho = m|\Psi|^{2}$ sourcing the gravitational field is itself determined by the quantum state.

The SNEs were originally proposed by Diósi and Penrose \cite{diosi,penrose96,penrose98} as a phenomenological modification of quantum mechanics intended to capture possible gravitational contributions to wavefunction reduction. Their motivation stems from the observation that coherent superpositions of macroscopically distinct quantum states would correspond to distinct spacetime geometries, a situation that is conceptually problematic within general relativity. In this context, the stationary SNEs were proposed as a candidate mechanism for gravity-induced suppression of macroscopic superpositions.

A second line of motivation arises from semiclassical gravity, in which the Einstein equations are sourced by the expectation value $\langle T_{\mu\nu}\rangle$ of the energy-momentum tensor, rather than a quantized gravitational field \cite{carlip}. In the weak-field, slow-motion limit, such semiclassical equations reduce precisely to the SNEs. Equivalent nonrelativistic limits can be obtained from systematic $1/c$ expansions of the Einstein--Klein--Gordon and Einstein--Dirac systems \cite{giulini2012}. Closely related equations also appear in astrophysical studies of self-gravitating systems (boson stars) \cite{schunk} and scalar-field models of dark matter \cite{guzman}.

The mathematical properties of the stationary SNEs have been examined extensively \cite{moroz,harrison,tod}, revealing the existence of self-bound states and characteristic scaling relations. Time-dependent numerical studies \cite{giulini2011,vanmeter,manfredi} have shown that gravitational self-interaction becomes dynamically relevant only above a critical mass scale $m_{c} \sim \hbar^{2/3} R^{-1/3} G^{-1/3}$, where $R$ denotes the spatial extent of the initial wave packet.

In more recent years, post-Newtonian corrections to the SNEs were derived by several authors~\cite{manfredi2015,brizuela2022relativistic}, as well as extensions to many-particle systems \cite{Giulini14}. Stationary solutions beyond spherical symmetry have also been constructed and their partial-wave content characterized~\cite{flores2025solitons}.  Optical analogues of the two-dimensional SNEs have been realized experimentally with nonlinear media and used to reproduce the time evolution of rotating boson stars~\cite{roger2016optical}. A recent extension of these experiments emulates post-Newtonian dynamics, exhibiting soliton solutions distinct from their Newtonian counterparts~\cite{paz2025nonlinear}. In astrophysical settings, the SNEs have been employed to model the linear stability of nonrelativistic boson stars~\cite{roque2023radial}, as well as fuzzy dark matter, an alternative to the usual cold dark matter consisting of ultralight bosons~\cite{manita2024soliton}.

Experimental validation of the predictions of the SNEs is also an active research area. 
A survey by Bassi and coworkers \cite{bassi2017} reviews the mechanisms by which classical and quantum gravitational fluctuations destroy quantum coherence, and the state of the art of tabletop experiments. 
Recent proposals include matter-wave interferometry with magnetically levitated superconducting microspheres, which could unambiguously falsify the Di\'osi--Penrose proposal~\cite{pino2017onchip}, and Stern--Gerlach interferometry of mesoscopic spheres, in which SN self-gravity induces a measurable phase shift between spin components~\cite{hatifi2020revealing}. Other, recent investigations exploit continuous-variable tomography of optomechanical oscillators~\cite{zhong2026tomography}. 

In summary, although they do not constitute a complete theory of quantum--gravitational interactions \cite{Anastopoulos,Bahrami}, the SNEs remain a mathematically tractable minimal model for investigating how Newtonian gravity might influence quantum dynamics, and serve as a useful theoretical framework for exploring semiclassical approximations of quantum gravity, gravitational decoherence mechanisms, and self-gravitating quantum systems such as boson stars and fuzzy dark matter.

%%%%%%%%%%%%%%%%%%%%%%%%%%%%%%%%%%%

\subsection{Born's rule}
% Born rule + Bohm--de Broglie

In quantum mechanics, the Born rule \cite{Born1926} stipulates that the probability to observe a quantum particle at a given position $\mathbf r$ is given by the square of the wavefunction $|\Psi(\mathbf r)|^2$. 
It is well known that, in certain nonlocal hidden-variable theories, the Born rule is not necessarily imposed as a fundamental condition. For instance, in Bohm--de Broglie mechanics \cite{Bohm1952}, if an ensemble of trajectories is initially distributed according to the Born rule, namely
$P(\mathbf{r},t=0)=|\Psi(\mathbf{r},t=0)|^2$, where $P(\mathbf{r},t)$ denotes the probability density of the position variable constructed from the esemble of trajectories, then this relation is preserved by the subsequent dynamics. Importantly, however, the equations of Bohm--de Broglie mechanics remain well-defined without assuming the Born rule as an initial condition. 
Then, if the Born rule is not satisfied at the initial instant, the particle distribution $P(\mathbf{r},t)$ and the quantum  distribution $|\Psi(\mathbf{r},t)|^2$ will generally remain distinct at later times.
Valentini \cite{valentini_signal-locality_1991} proposed interpreting the Born rule as the quantum analogue of thermal equilibrium in classical statistical mechanics. In this interpretation, the Born rule represents ``quantum equilibrium" and the non-equilibrium path to such equilibrium is termed ``subquantum dynamics".
From a dynamical point of view, non-equilibrium distributions will typically converge to quantum equilibrium.
Earlier numerical simulations \cite{valentini2005dynamical} showed that relaxation to equilibrium is indeed observed, provided some coarse graining procedure is applied.

% Experiments
Recent experimental proposals to test the Born rule propose using helium photoionization and attosecond dynamics for a Sorkin-type test \cite{Foerderer2025}. Valentini and Varma proposed tests using spin and polarization probabilities in high-energy collisions, potentially probing quantum processes on time scales of order $10^{-25}\,\mathrm{s}$ \cite{Valentini2025}. 

In a cosmological setting, the possibility that signatures of such subquantum dynamics could have survived in the early universe and left observable imprints on primordial fluctuations has also been investigated \cite{Valentini2010,Underwood2015}. 
The expansion of the universe may have amplified or otherwise preserved these signatures, allowing deviations from quantum equilibrium to manifest in the primordial perturbation spectrum and, consequently, in the cosmic microwave background. In particular, such effects could lead to a suppression of power at large angular scales
\cite{Valentini2010,ColinValentini2015,ColinValentini2016}.

\subsection{Nelson's stochastic dynamics}
% Nelson

An alternative framework for investigating relaxation toward quantum equilibrium is provided by Nelson's stochastic mechanics \cite{nelson_derivation_1966,bacciagaluppi1999nelsonian,Beyer2021}. Nelson's dynamics shares important features with  Bohm--de Broglie mechanics, but differs in that the particle dynamics is stochastic rather than deterministic. Related stochastic approaches to quantum dynamics have also been considered by Bohm and Hiley \cite{BohmHiley1989}, Peruzzi and Rimini \cite{peruzzi1996}, and Bohm and Vigier \cite{BohmVigier1954}.

Nelson's theory reproduces the predictions of standard quantum mechanics when the Born rule is satisfied initially. If the initial distribution instead differs from the Born distribution, the stochastic dynamics can drive $P(\mathbf{r},t)$ toward the quantum equilibrium value $|\Psi({\mathbf r},t)|^2$ without requiring an externally imposed coarse-graining procedure. The stochastic character of the dynamics thus provides a natural mechanism for relaxation toward quantum equilibrium and makes Nelson's framework particularly well suited for investigating possible subquantum dynamics and the dynamical emergence of the Born rule. In an previous work, we studied numerically the relaxation toward quantum equilibrium for a double-slit setup, a harmonic oscillator, and a particle in a uniform gravitational field \cite{Hardel2023}. Earlier work had already investigated quantum-mechanical distributions as attractors for Nelson processes \cite{Petroni1995}.

%\medskip
\subsection{Outline of the present work}

In this work, we study the validity of the Born rule for semiclassical gravity scenarios modeled by the SNEs. In Sec. \ref{sec:variational}, using a variational method and assuming a Gaussian ansatz for the wavefunction, we obtain an approximate, but accurate, solution to the SNEs \cite{manfredi}. This approach provides all the information needed to construct the deterministic part of the velocity entering Nelson's model (Sec. \ref{sec:Nelson}), while the stochastic term is a white noise characterized by a diffusion coefficient $D_Q = \hbar/2m$. A Fokker--Planck equation, equivalent to Nelson's stochastic process, is also derived.
In Sec. \ref{sec:tests}ed we use Nelson's approach to assess the validity of the Born rule for a semiclassical  gravity scenario described by the SNEs with spherical symmetry. Through numerical simulations and analytical considerations, we establish under which initial conditions the Born rule is approach for long times.
Conclusions and perspectives are presented in Sec. \ref{sec:conclusion}.

\section{Variational solution to the Schrödinger-Newton equations}\label{sec:variational}

To obtain an approximate solution to the SNEs, we adopt the variational method detailed in Ref. \cite{manfredi}, which we briefly sketch here.

We first rewrite the SNEs \eqref{eq:schrodinger}-\eqref{eq:newton} in dimensionless form. Length, energy, and time are measured in units of  $a_G=\hbar^2/(Gm^3)$, $E_G=m^5G^2/\hbar^2$, and $t_G=\hbar/E_G$, respectively. To preserve its norm, the wavefunction is laso normalized to $a_G^{-3/2}$.
The dimensionless SNEs become
\begin{eqnarray}
i \frac{\partial \Psi}{\partial t} &=& -\frac{1}{2} \Delta \Psi +V({\mathbf r},t) \Psi,
\label{schrodinger_norm} \\
\Delta V &=& 4\pi |\Psi|^2,
\label{poisson_norm}
\end{eqnarray}
with $\int |\Psi|^2 d{\mathbf r}=1$. These equations contain no parameters, so the evolution is fully determined by the initial condition. For a spherically symmetric Gaussian profile, the only relevant dimensionless parameter is its initial width in units of $a_G$. All forthcoming numerical results will  be expressed in these normalized units.

We rewrite Eqs. (\ref{schrodinger_norm})--(\ref{poisson_norm}) in hydrodynamic form using the Madelung transformation $\Psi=\sqrt{\rho}\exp(iS)$. The resulting continuity and momentum equations are
\begin{eqnarray}
\label{continuity}
\frac{\partial \rho}{\partial t} &+& \nabla\cdot (\rho{\mathbf u}) = 0 ,\\
\label{momentum}
\frac{\partial{\mathbf u}}{\partial t} &+& {\mathbf u}\cdot\nabla{\mathbf u}  = - \nabla V + \frac{1}{2} \,\nabla
\left(\frac{\nabla^2 \sqrt{\rho}}{\sqrt{\rho}} \right) ,
\end{eqnarray}
with ${\mathbf u}=\nabla S$. Together with Eq. (\ref{poisson_norm}), these equations follow from the Lagrangian density 
\be
{\cal L}(\rho,S,V) = \frac{\rho}{2}\left(\nabla S\right)^2 + \rho \frac{\partial S}{\partial t} +  \frac{\left(\nabla \rho\right)^2}{8\rho}
+\frac{(\nabla V)^2}{8\pi} +\rho V .
\label{lag_density}
\ee
We now seek a reduced description in terms of macroscopic variables. Assuming spherical symmetry and a Gaussian density profile
\begin{equation}
\label{gauss}
\rho({\mathbf r},t) = \frac{1}{\pi^{3/2} R^3(t)}\,\exp\left(- \frac{r^2}{R^2(t)}\right) ,
\end{equation}
where $r=|{\mathbf r}|$, Poisson's equation has the exact solution
\begin{equation}
V({\mathbf r},t) = -\frac{1}{r} \,{\rm erf}\left(\frac{r}{R(t)}\right) ,
\label{potential}
\end{equation}
while the continuity equation is solved by ${\mathbf u}({\mathbf r},t)=(\dot R/R){\mathbf r}$, corresponding to $S=(\dot R/2R)r^2$, where the dot denotes time differentiation.

Substituting these expressions into Eq. (\ref{lag_density}) and integrating over space yields
\be
L(R,\dot{R}) = -{2\over3}\int{\cal L}\,d{\bf r} = \frac{\dot{R}^2}{2} - \frac{1}{2R^2} + \frac{C}{R},
\label{lag_au}
\ee
where $C=2/(3\sqrt{2\pi})$ and the multiplying factor $-2/3$ was chosen for notation convenience. The Euler--Lagrange equation for $L(R,\dot{R})$ then yields
\be
\frac{d^2 R}{d t^2} = \frac{1}{R^3} - \frac{C}{R^2}.
\label{eq:motion}
\ee

\begin{figure}[ht]
\centering
\includegraphics[width=0.45\textwidth]{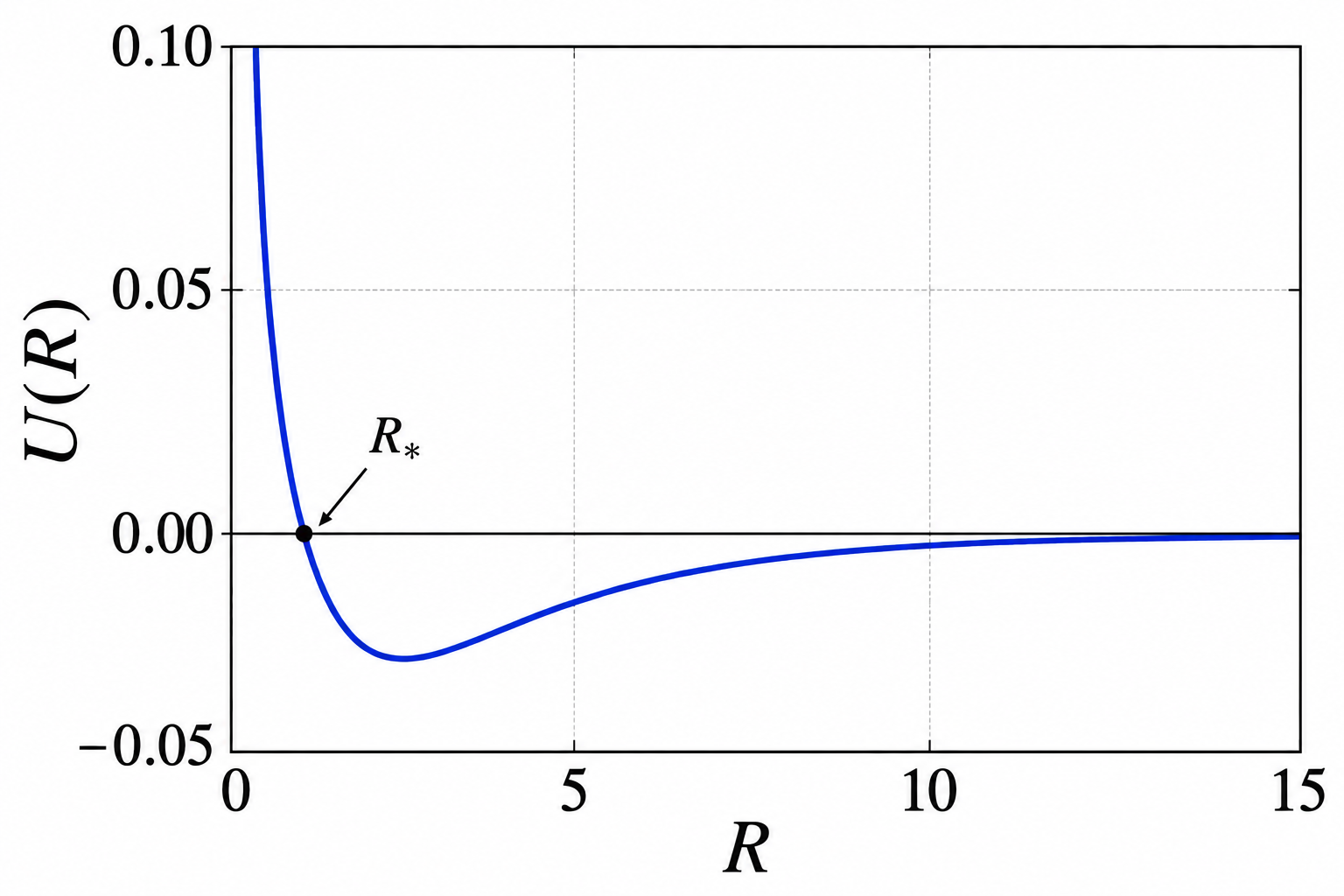}
\caption{Profile of the effective potential $U(R)=1/(2R^2) - C/R$.} \label{fig:poten}
\end{figure}

Equation (\ref{eq:motion}) is equivalent to the motion of a point particle in the effective potential $U(R)=1/(2R^2)-C/R$ (see Fig. \ref{fig:poten}). The first term is repulsive and represents quantum kinetic energy, while the second represents attractive self-gravity.
The pseudo-potential $U(R)$ diverges as $R\to 0$ and vanishes as $R^{-1}$ for $R\to\infty$. It crosses zero at $R_{\displaystyle *}=1/2C \approx 1.88$ and has a single minimum at $2R_{\displaystyle *}$.
As shown in \cite{manfredi}, this approximate Gaussian ansatz fares rather well in approximating the exact solution  of the SNEs, both for the ground state and for time-dependent evolutions.

Depending on its initial condition, and always assuming zero initial velocity $\dot R(0)=0$, one obtains that $R(t)$ behaves as  $R(t) \sim  t$ for $R(0)<R_{\displaystyle *}$, is bounded for $R(0)> R_{\displaystyle *}$, and behaves as $R(t) \sim  t^{2/3}$ for $R(0)=R_{\displaystyle *}$ \cite{manfredi}.
Interestingly, in a cosmological setting, these regimes correspond respectively to an open, closed, and Einstein-de Sitter universe.
In the following, we will show that they also display different behaviours as to their convergence to the Born rule.

%%%%%%%%%%%%%%%%%%%%%%%%%%%%%%%%%%%%%%%%%%%%%%%%%%%%%%%%%%%%%%%%%%%

\section{Nelson's stochastic quantization}\label{sec:Nelson}

Nelson's stochastic formulation of quantum mechanics \citep{nelson_derivation_1966,bacciagaluppi1999nelsonian} is similar to the Bohm--de Broglie theory in that it provides an evolution equation for the particle's trajectory. In contrast to Bohmian mechanics, however, in Nelson's mechanics the particle position evolves according to the stochastic differential equation
\begin{equation}\label{Nelson}
\dd {\mathbf r}(t)= {\mathbf b}({\mathbf r}(t),t)\,\dd t+\dd {\mathbf W}(t),
\end{equation}
where ${\mathbf b}({\mathbf r}(t),t)$ represents the drift velocity field and ${\mathbf W}(t)$ is a Wiener process with vanishing mean
$\langle\dd {\mathbf W}\rangle=0$ and variances
$\langle \dd W_{i}\,\dd W_{j}\rangle = 2 \delta_{ij} D_Q $, where $\delta_{ij}$ is the Kronecker delta and $(i,j)$ represent the three spatial directions. 

The drift field appearing in Eq.~\eqref{Nelson} can be decomposed into a phase-dependent contribution and an amplitude-dependent contribution according to
\begin{equation}\label{b_Nelson}
{\mathbf b}({\mathbf r},t) = 2D_Q \nabla S({\mathbf r},t)
+ D_Q\nabla\ln \rho({\mathbf r},t).
\end{equation}
The first term corresponds to the drift velocity and is proportional to the spatial gradient of the phase. It therefore coincides with the Bohm--de Broglie guidance velocity. The second contribution, commonly referred to as the osmotic velocity, depends instead on the spatial variation of the wavefunction amplitude.

The stochastic dynamics can equivalently be formulated at the level of the probability density. In particular, the Langevin equation~\eqref{Nelson} leads to the corresponding Fokker--Planck equation
\begin{equation}\label{eq:FokkerPlanck}
\frac{\partial P}{\partial t}
+ \nabla \cdot \big({\mathbf b}({\mathbf r},t)P\big)
= D_Q \,\Delta P,
\end{equation}
which describes the temporal evolution of the probability density $P({\mathbf r},t)$.

A crucial property of Nelson's formulation is that, provided the initial particle distribution satisfies
\[
P({\mathbf r},0)=|\Psi({\mathbf r},0)|^2 \equiv \rho({\mathbf r},0),
\]
the subsequent evolution reproduces the predictions of standard quantum mechanics. In this sense, Nelson's theory is empirically equivalent to the conventional Schr\"odinger description, like the Bohm--de Broglie approach. The two frameworks nevertheless differ fundamentally in their description of the underlying trajectories: whereas Bohm--de Broglie mechanics assigns deterministic trajectories to particles, Nelsonian mechanics incorporates an intrinsically stochastic component.

One should emphasize, however, that the stochastic character of the trajectories does not imply an irreversibility of the complete dynamics. Nelson's theory is time-reversal invariant \cite{nelson_derivation_1966}, a property that is essential for maintaining its equivalence with the time-dependent Schr\"odinger equation. This reversibility can be made explicit at the level of the probability distribution. Indeed, upon inserting the decomposition~\eqref{b_Nelson} into the Fokker--Planck equation~\eqref{eq:FokkerPlanck}, the diffusive contribution generated by the Wiener process is exactly compensated by the osmotic component of the velocity field. The resulting evolution therefore possesses the same time-reversal structure as the underlying quantum dynamics.

% application to SNEs
In order to apply Nelson's theory to the reduced SNEs of Sec. \ref{sec:variational}, we should focus on the evolution in spherical geometry, where $P$ is a function of the radius $r=|{\mathbf r}|$ only, plus time. We define the spherically-symmetric particle density: $\widetilde{P}(r,t) = 4\pi r^2 P(r,t)$, normalized as $\int_0^{\infty} \widetilde{P}(r,t) \,\dd r = 1$.
By so doing, the Fokker--Planck equation \eqref{eq:FokkerPlanck} becomes
\begin{equation}\label{eq:FokkerPlanck-spher}
\frac{\partial \widetilde{P}}{\partial t}
+ \frac{\partial}{\partial r} \left[ \left(b(r,t) +\frac{2D_Q}{r}\right) \widetilde{P} \right]
= D_Q \,\frac{\partial^2 \widetilde{P}}{\partial r^2},
\end{equation}
with the corresponding stochastic process 
\begin{equation}\label{eq:Nelson-spher}
\dd  r=  \big(b(r,t) + 2 D_Q/r\big) \,\dd t+\dd  W.
\end{equation}

We now have all the elements to compute the drift velocity $b(r,t)$  as defined in Eq. \eqref{b_Nelson}. Employing the Gaussian ansatz detailed in Sec. \ref{sec:variational}, with the phase $S=(\dot R/2R)r^2$ and the density $\rho$ given by Eq. \eqref{gauss}, we obtain

\be
b(r, t) = \left(\frac{\dot R}{R} - \frac{2 D_Q}{R^2}\right)r(t).
\label{eq:b-spher}
\ee
We recall that, in our units, $D_Q=\hbar/2m=1/2$.

The stochastic process \eqref{eq:Nelson-spher}--\eqref{eq:b-spher}, together with the evolution of $R(t)$ given by Eq. \eqref{eq:motion}, constitutes our model to assess the validity of the Born rule within a Newtonian semiclassical gravity setting. 
In particular, we will set initially $P(r,0) \neq |\Psi({ r},0)|^2$ and examine whether the Born rules emerges at later times.

%%%%%%%%%%%%%%%%%%%%%%%%%%%%%%%%%%

\section{Tests of the Born rule}\label{sec:tests}

\subsection{Analytical results}\label{sec:analytical}

Using the Fokker--Planck equation \eqref{eq:FokkerPlanck-spher}, it is easy to compute the evolution equation of the second moment $\langle r^2 \rangle(t)$:
\be\label{eq:r2}
\frac{\dd \langle r^2 \rangle}{\dd t} = \left(2\frac{\dot R}{R} - \frac{4D_Q}{R^2} \right) \langle r^2 \rangle + 6 D_Q .
\ee
If the particle distribution ${\widetilde P}(r,t)$ is Gaussian, 
\begin{equation}
\label{gauss-part}
{\widetilde P}(r,t) = \frac{4\pi r^2}{\pi^{3/2} R_p^3(t)}\,\exp\left(- \frac{r^2}{R_p^2(t)}\right) ,
\end{equation}
then $\langle r^2 \rangle = {3 \over 2} R_p^2$. The width $R_p(t)$
is directly comparable to the width $R(t)$ computed from the spherically-symmetric wavefunction density ${\widetilde\rho}(r,t) \equiv 4\pi r^2 \rho(r,t)$ using Eq. \eqref{gauss}.
The width $R_p(t)$ obeys the equation
\be\label{eq:rp}
\frac{\dd R_p}{\dd t} = \left(\frac{\dot R}{R} - \frac{2D_Q}{R^2} \right)R_p + 2 D_Q /R_p,
\ee
which is fully determined once $R(t)$ is known from Eq. \eqref{eq:motion}. Note that the evolution equation for $R_p(t)$ is of first order in time, while that for $R(t)$ is second order. This is a result of Nelson's stochastic process being of first order, i.e., overdamped.

If the particles are initially distributed according to the Born rule, then $R_p(t)=R(t)$ for all times.
In order to study the approach to the Born rule, we define the quantity
\be\label{eq:z}
z(t) = \frac{R_p^2}{R^2} -1.
\ee
The Born rule is satisfied when $z=0$, while for $z>0$ (resp., $z<0$) the particle distribution has a wider (resp., narrower) spread than the quantum distribution.
Employing Eqs. \eqref{eq:rp} and \eqref{eq:motion}, we obtain that $z(t)$ obeys the following linear equation:
\be\label{eq:z-evol}
\frac{\dd z}{\dd t} = -\frac{4D_Q}{R^2(t)}\, z ,
\ee
which has the general solution
\be\label{eq:z-solution}
z(t) = z(0)\, \exp\left(- \int_0^t\frac{4D_Q}{R^2(s)}\,\dd s\right).
\ee

Hence, the fate of $z(t)$ for large times is determined by the evolution of the wavefunction, which in turn is uniquely determined by $R(t)$. If the integral in Eq. \eqref{eq:z-solution} diverges, then $z(t) \rightarrow 0$ as $t \to \infty$, and the Born rule is asymptotically satisfied; if the integral remains finite, then also $z(t)$ will remain finite, and the Born rule is violated for all times~\footnote{The integral in Eq. \eqref{eq:z-solution} diverges also for unbound solutions, provided that $R(t)$ grows slower than $t^{1/2}$. Interestingly, this is the expansion law of a universe dominated by radiation.}.
In terms of our SNEs, this means that the Born rule is asymptotically satisfied for bound solutions of Eq. \eqref{eq:motion}, i.e., solutions for which $R(0) > R_{\displaystyle *}$, whereas it is violated for unbound solutions $R(0) \le R_{\displaystyle *}$.

\subsection{Numerical results}\label{sec:numerical}

These findings are corroborated by simulations of the stochastic Nelson model. 
All numerical results are expressed in the normalized units defined in Sec. \ref{sec:variational}

\begin{figure}[ht]
\centering
\includegraphics[width=0.5\textwidth]{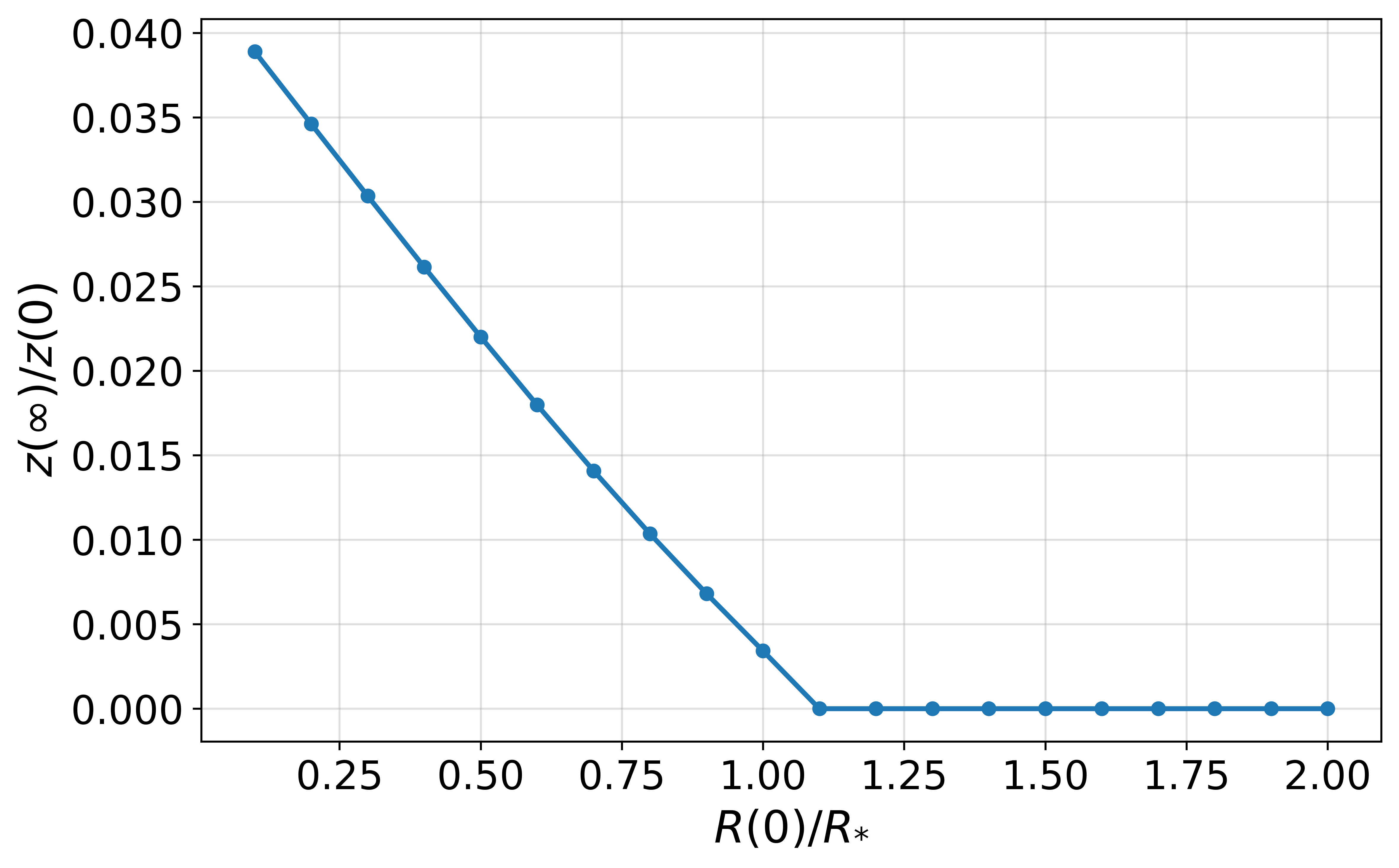}
\caption{Late-time values of $z(t)$, normalized to $z(0)$, as a function of the initial wave-packet width $R(0)$, normalized to $R_{\displaystyle *}$. The results are obtained from the analytical solution of Eq. \eqref{eq:z-evol}, for different evolutions of $R(t)$ corresponding to bound or unbound states.}
\label{fig:z}
\end{figure}

Figure \ref{fig:z} shows the late-time value of $z(t)$ as a function of the initial value $R(0)$, normalized to $R_{\displaystyle *}$. As expected, $z(t \to \infty)$ remains finite for $R(0) \le R_{\displaystyle *}$, while it vanishes for $R(0) > R_{\displaystyle *}$.

\begin{figure}[ht]
\centering
\includegraphics[width=0.8\textwidth]{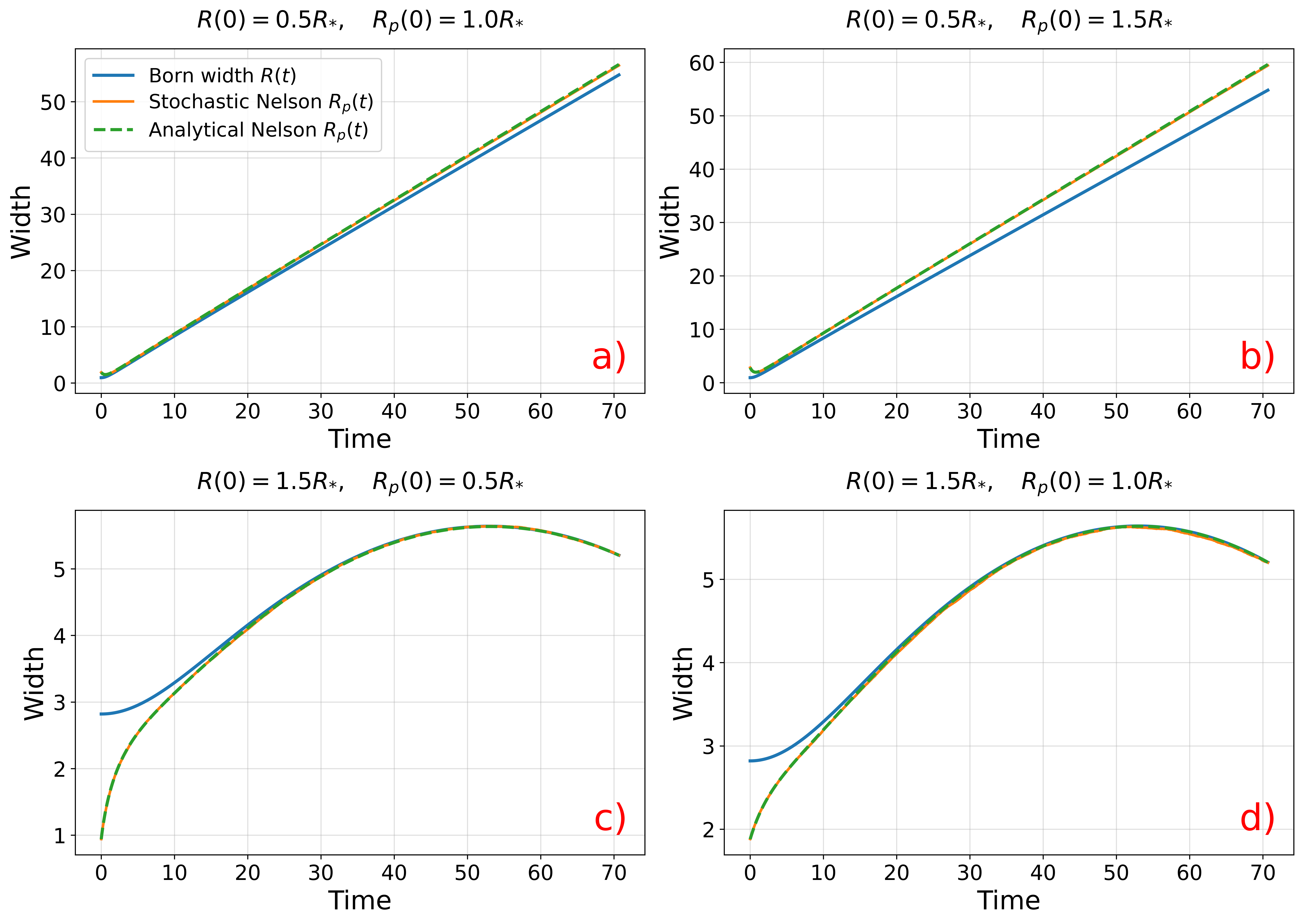}
\caption{Evolutions of $R(t)$ from Eq. \eqref{eq:motion} (blue lines), $R_p(t)$ from Eq. \eqref{eq:rp} (dashed green line), and $R_p(t)$ computed from the stochastic Nelson equation \eqref{eq:Nelson-spher} (orange lines). The top panels (a) and (b) correspond to unbound cases with $R(0)= 0.5 R_{\displaystyle *}$, and two different intial values of $R_p(0)$. The bottom panels (c) and (d) correspond to bound cases with $R(0)= 1.5 R_{\displaystyle *}$, and two different initial values of $R_p(0)$.}
\label{fig:Rp}
\end{figure}

In Fig. \ref{fig:Rp}, we show the evolution of $R(t)$ and $R_p(t)$ for different values of $R(0)$, yielding bound or unbound solutions. It clearly appears that these two quantities converge to each other in the bound cases [panels (c) and (d)], while they diverge in the unbound cases [panels (a) and (b)], in accordance with the theory.

\begin{figure}[ht]
\centering
\includegraphics[width=0.5\textwidth]{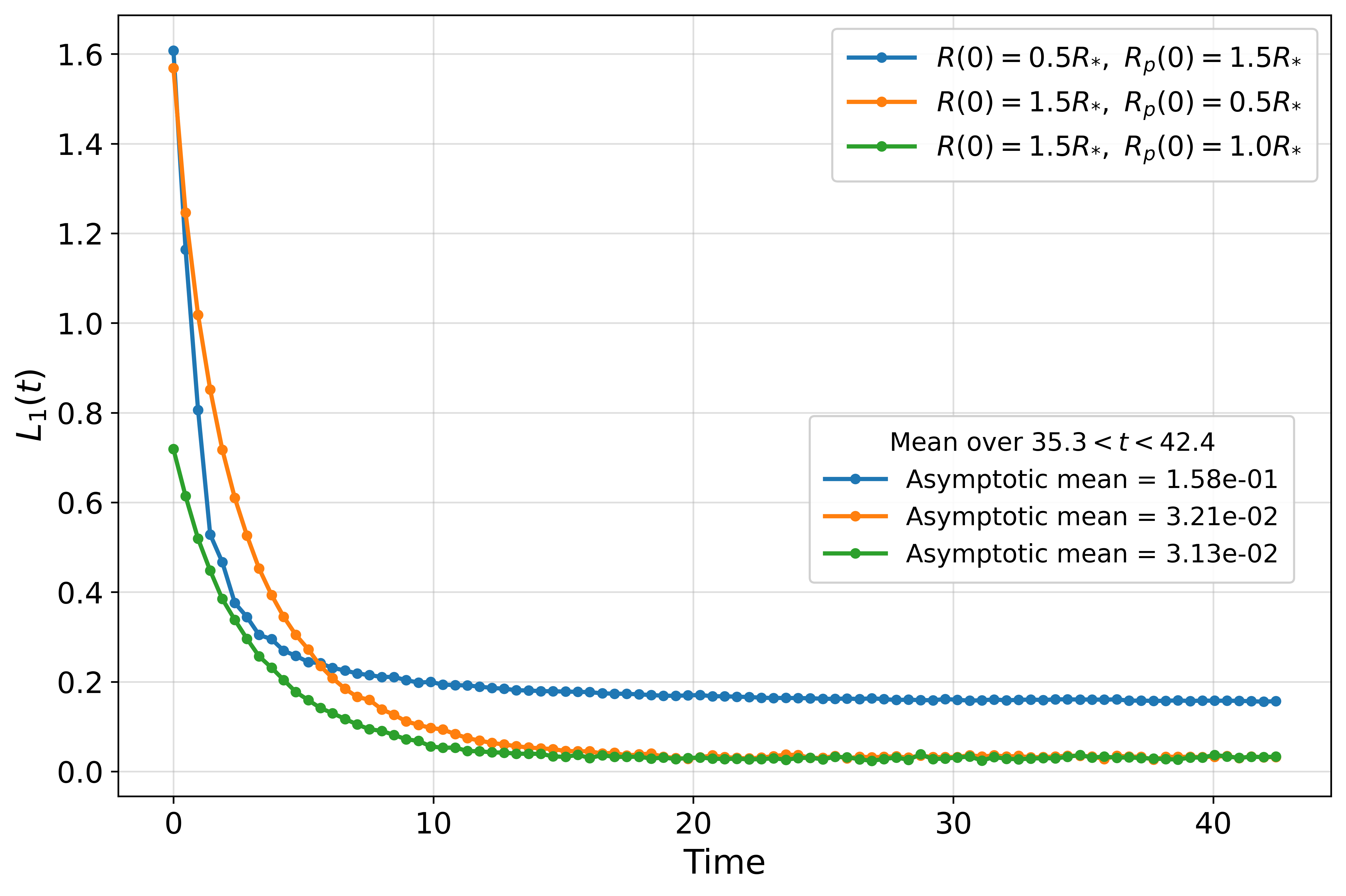}
\caption{$L_1$ norm for one unbound evolution with $R(0) = 0.5R_{\displaystyle *}$ (blue dots) and two bound evolutions with $R(0) =1.5R_{\displaystyle *}$. The lines connect the dots, which were obtained by simulating the stochastic Nelson process. The lower inset shows the mean of the $L_1$ norm computed between $t=35.3$ and $t=42.4$. }
\label{fig:L1}
\end{figure}

So far, we have established the convergence, or otherwise, to the Born rule by checking the match between the widths $R(t)$ and $R_p(t)$ of the probability densities. However, this does not guarantee that the densities themselves, ${\widetilde \rho}(r,t)$ and ${\widetilde P}(r,t)$, are identical. Here, we use a different measure, namely the $L_1$ norm, defined as
\be
L_1(t) = \int_0^\infty |{\widetilde P}(r,t) - {\widetilde \rho}(r,t)|\, dr,
\label{eq:L1}
\ee
which is shown in Fig. \ref{fig:L1}. 
Contrarily to $z(t)$, the $L_1$ norm is a global measure of the match between the two distributions ${\widetilde \rho}(r,t) \equiv 4\pi r^2 |\Psi|^2$ and ${\widetilde P}(r,t)$.
For the case with $R(0) < R_{\displaystyle *}$, the $L_1$ norm remains finite, signalling a departure from the Born rule for log times. For $R(0) > R_{\displaystyle *}$ the asymptotic value of $L_1$ is effectively zero, in accordance with the results based on the widths of the distributions. The asymptotic value of $L_1$ is not exactly zero due to the finite number of simulated trajectories.

\begin{figure}[ht]
\centering
\includegraphics[width=0.85\textwidth]{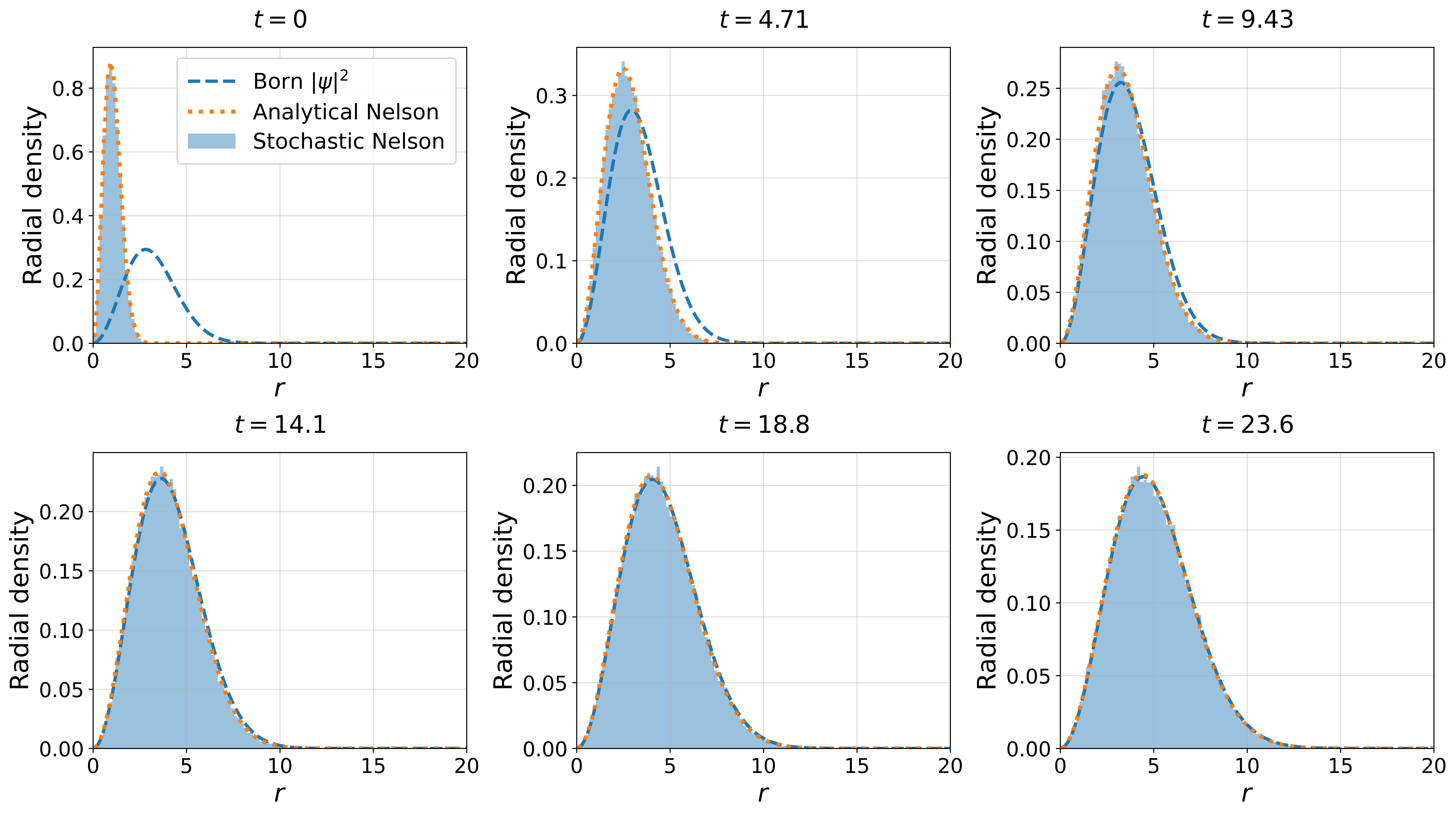}
\caption{Evolution of the probability distributions for a bound case with $R(0) = 1.5 R_{\displaystyle *}$ and $R_p(0) = 0.5 R_{\displaystyle *}$. We show several snapshots, at different times, of the wavefunction density $\widetilde{\rho}(r,t) \equiv 4\pi r^2 |\Psi|^2$ (dashed blue line), particle density $\widetilde{P}(r,t)$ from Nelson's stochastic process (blue histograms), and Gaussian particle density \eqref{gauss-part}, with $R_p(t)$  computed from Eq. \eqref{eq:rp} (dot-dashed orange line).
}
\label{fig:bound}
\end{figure}

\begin{figure}[ht]
\centering
\includegraphics[width=0.85\textwidth]{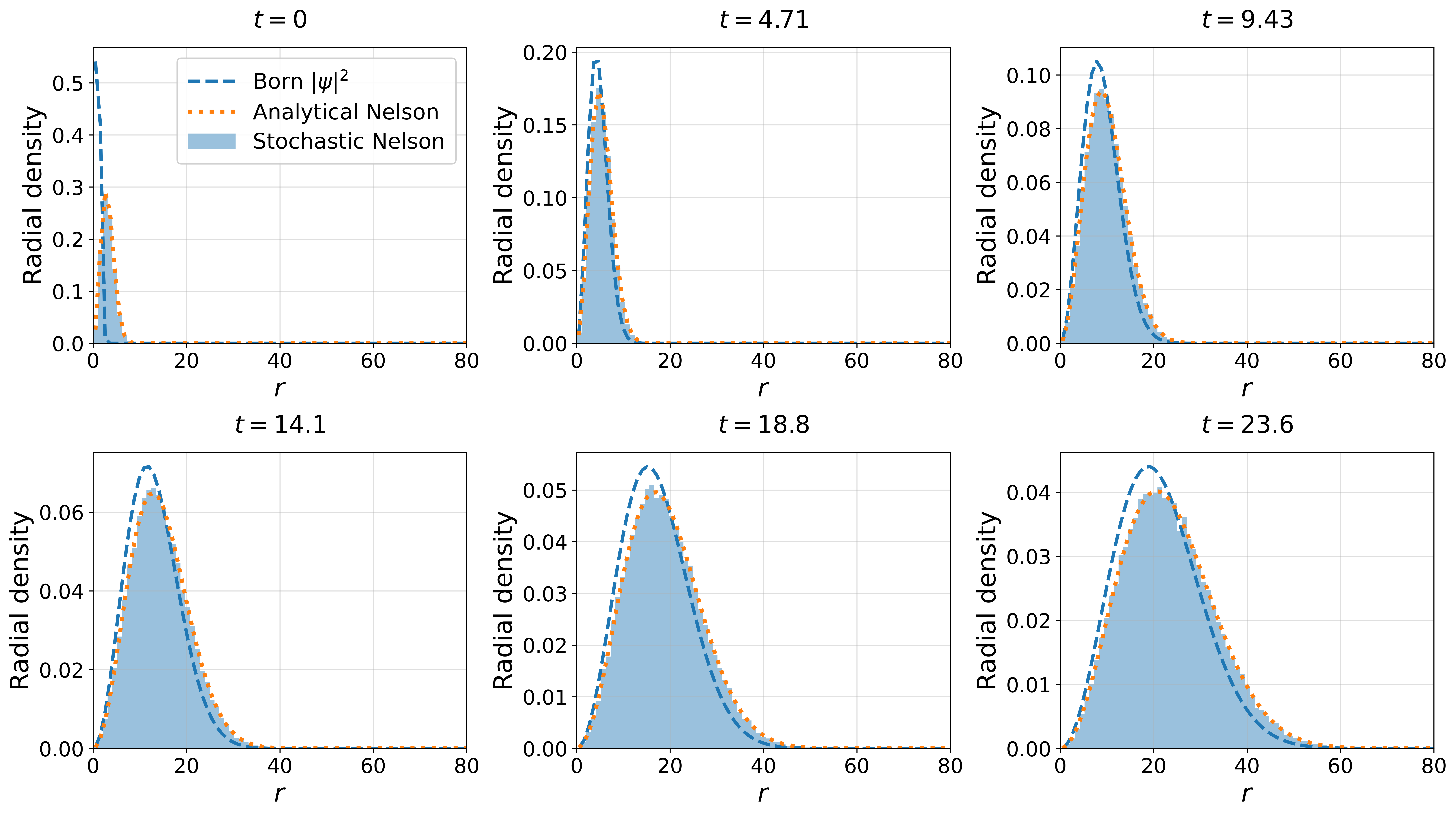}
\caption{Same as Fig. \ref{fig:bound} for an unbound case with $R(0) = 0.5 R_{\displaystyle *}$ and $R_p(0) = 1.5 R_{\displaystyle *}$.}
\label{fig:unbound}
\end{figure}

We next show the full distributions for a bound case (Fig. \ref{fig:bound}, $R(0) =0.5R_{\displaystyle *}$) and an unbound case (Fig. \ref{fig:unbound}, $R(0) =1.5R_{\displaystyle *}$). 
The quantum distribution ${\widetilde \rho}(r,t)$ (dashed blue line) is Gaussian by construction of the model.
The particle distribution ${\widetilde P}(r,t)$ (blue histograms) is also initially Gaussian and remains so throughout the evolution, as evidenced by the fact that the histograms follow closely the analytical Gaussian curve with width $R_p(t)$ (dot-dashed orange line). 
In the bound case (Fig. \ref{fig:bound}) the convergence to quantum equilibrium is clearly observed for the full particle distribution. in contrast, violation of the Born rule occurs in the unbound case (Fig. \ref{fig:unbound}).

\subsection{Fokker--Planck equation in comoving co-ordinates}\label{sec:comoving}
We consider the Fokker--Planck equation \eqref{eq:FokkerPlanck-spher} for the particle density $\widetilde{P}(r,t)$, with the drift velocity \eqref{eq:b-spher}. 
Following a standard procedure in cosmology, we rescale the space and time variables as follows:
\be\label{eq:rescaling}
r = R(t) \,q, \qquad \dd t = \frac{R^2(t)}{D_Q}\, \dd\tau,
\ee
where $(q,\tau)$ represent the new space-time variables. These are known in cosmology as supercomoving co-ordinates \cite{Martel1998}.
We also transform the probability density $\widetilde{P}(r,t) = {\cal P}(q,\tau) / R(t)$, so that $\int_0^\infty \widetilde{P} \dd r = \int_0^\infty {\cal P} \dd q = 1$. Note that $q$, $\tau$, and ${\cal P}$ are dimensionless. 

After some algebra, the Fokker--Planck equation \eqref{eq:FokkerPlanck-spher} is transformed to the following
\be
\label{eq-FP-scaled}
\frac{\partial {\cal P}}{\partial \tau}
+ \frac{\partial}{\partial q} \left[ \left(-2 q  + \frac{2}{q}\right){\cal P}\right] 
=  \,\frac{\partial^2 {\cal P}}{\partial q^2}\,,
\ee
where, importantly, the coefficients are all time-independent.

The equilibrium solution of the above Fokker--Planck equation reads as
\be
\label{eq:FP-equilibrium}
{\cal P}_{\infty}(q) = \frac{4\pi q^2}{\pi^{3/2}}\, e^{-q^2}.
\ee
Transforming back to the physical co-ordinates $(r,t)$, one obtains:
\be
\widetilde{P}_{\infty}(r,t) = \frac{4\pi r^2}{\pi^{3/2} R^3(t)}\, e^{-r^2/R^2(t)} \,,
\ee
which is identical to the spherically-symmetric wavefunction density $\widetilde{\rho}(r,t) \equiv 4\pi r^2 \rho(r,t)$, see Eq. \eqref{gauss}.
Since the  equilibrium solution \eqref{eq:FP-equilibrium} is always attained for $\tau \to \infty$, it would appear that the particle density $\widetilde{P}$ always tends to the wavefunction density, and thus that the Born rule is always satisfied asymptotically.

However, there is a caveat. The rescaled time $\tau$ does not always tend to infinity when $t \to \infty$. Indeed
\be
\tau(t) = \int_0^t \frac{D_Q}{R^2(s)} \,\dd s
\ee 
goes to infinity only if $R(t)$ is bound or grows slower than $t^{1/2}$. If it grows faster, e.g. $R(t) \sim t$, as in the unbound solutions of the SNEs, then $\tau(t \to \infty)$ tends to a finite value.  In other words, the rescaled time is constrained on a finite interval $[0, \tau(\infty)]$. In that case, the initial distribution will never reach the equilibrium ${\cal P}_{\infty}(q)$, and the Born rule will not be statisfied asymptotically.
Thus, we have recovered the same results as in Sec. \ref{sec:analytical}, which were confirmed numerically in Sec. \eqref{sec:numerical}.

We note that the comoving Fokker--Planck approach goes further than the preceding results, as it does not assume the particle distribution to be a spherically-symmetric Gaussian: any arbitrary initial distribution converges, in comoving space, to the stationary solution \eqref{eq:FP-equilibrium}.

\section{Conclusion}\label{sec:conclusion}

In this work, we have investigated the emergence of the Born rule in a semiclassical-gravity setting described by the nonlinear Schrödinger--Newton equations. By combining a Gaussian variational description of the wavefunction with Nelson's stochastic formulation of quantum mechanics, we have been able to follow the evolution of particle distributions that do notinitially  respect the Born rule. 

This approach is particularly relevant in light of the broader programme initiated by Valentini \cite{Valentini1991}, in which the Born rule is interpreted as a state of ``quantum equilibrium" rather than as a fundamental probabilistic postulate.
In Bohmian or Nelsonian mechanics, an ensemble initially satisfying the Born rule, $P({\mathbf r})=|\Psi|^2({\mathbf r})$, remains in quantum equilibrium for all later times. Non-equilibrium distributions will, most of the time, relax to quantum equilibrium \cite{Valentini1991}, although cases where this does not happen on a significantly long timescale have been pointed out \cite{Hardel2023}.

Here, our main result is that relaxation toward the Born distribution is controlled by the long-time behaviour of the Schr\"odinger--Newton wave packet. When the wave packet remains confined, or expands sufficiently slowly, the stochastic dynamics has enough time to relax to quantum equilibrium. In contrast, when the wave packet expands sufficiently rapidly, the available relaxation time is finite, and an initially non-equilibrium distribution can retain some memory of its initial non-Born state.
Numerical simulations of the stochastic Nelson process have confirmed this picture.

In a cosmological context, Valentini and coworkers have argued that relaxation to quantum equilibrium might have been incomplete in the early universe, so that quantum non-equilibrium could survive and potentially leave observable signatures in the primordial fluctuation spectrum and the cosmic microwave background \cite{Valentini1991,Valentini2010,Underwood2015,ColinValentini2015,ColinValentini2016}. 
Our results complement this suggestion from a different perspective. Rather than considering quantum non-equilibrium in an expanding cosmological background, we have shown that even the simple dynamics of a self-gravitating Schr\"odinger--Newton system can determine whether quantum equilibrium is dynamically attainable. 
In other words, if quantum non-equilibrium was indeed present in the early universe, subsequent cosmic expansion may have inhibited its relaxation towards the Born rule and thereby preserved, or potentially even amplified, its observable signatures.

Several extensions of the present work are worth pursuing. First, the Gaussian variational approximation could be replaced by numerical solutions of the full time-dependent SNEs, allowing the relaxation mechanism to be studied beyond the restricted family of Gaussian states. Second, the assumption of spherical symmetry could be relaxed in order to investigate anisotropic configurations and more general initial states. Finally, incorporating a genuinely cosmological expansion, and eventually a relativistic post-Newtonian treatment (see, for instance, \cite{manfredi2015}), would provide a natural bridge between the present semiclassical model and a realistic cosmological scenario.

%%%%%%%%%%%%%%%%%%%%%%%%%%%%%%%%%%%%%%%%%%%%%%%%%%%%%%%%%%

\bibliography{SNEbiblio,biblionelson}

\end{document}